\documentstyle[11pt,newpasp]{article}
 
\def\deg{$^{\circ}\,$}
\def\solm{M$_{\odot}\,$}

\def\etal{{\it et~al.\ }}
\def\eg{{\it e.g.\ }}

\begin{document}

\title{Gas kinematics from spectroscopy with a wide slit: \\
detecting nuclear black holes}

\author{Witold Maciejewski and James Binney} 
\affil{Theoretical Physics, University of Oxford}

\begin{abstract}
Motivated by STIS observations of more than 50 nearby galactic nuclei,
we consider long-slit emission-line spectra when the slit is wider
than the instrumental PSF, and the target has arbitrarily large velocity
gradients. The finite width of the slit generates complex patterns in the 
spectra that can be misinterpreted as coming from various physically 
distinct nuclear components, but when interpreted correctly, they can 
have considerable diagnostic power. For 
a thin disk in circular motion around a central galactic
black hole (BH), a characteristic artifact occurs in the spectrum at the 
outer edge of the BH's sphere of 
influence. It betrays the presence of a BH, and allows us 
to develop a new method for estimating its mass, which gives higher 
sensitivity to BH detection than traditional methods.
\end{abstract}

\section{Introduction}
Observations with long-slit spectrographs often use a slit 
that is wider than the FWHM of the instrumental point-spread-function (PSF). 
This practice is conventionally considered to enhance the 
signal-to-noise ratio (S/N) of the data at the price of what may be 
an insignificant loss in velocity resolution. However, when the 
target has steep velocity gradients, the use of a wide slit can have 
more subtle effects, because the position and velocity information become 
entangled along the dispersion direction. If these
effects are not recognized in the data, a misleading impression of the
structure of the target can be inferred.  When the effects of
a wide slit are recognized, however, they can increase the diagnostic power
of the spectrum over that of a narrow-slit spectrum of equal S/N.

This study has been motivated 
by our participation in STIS observations of more than 50 nearby galactic 
nuclei (see Marconi \etal 2000, this volume), therefore we focus on the 
effect that a wide slit has on the emission-line spectra from gaseous disks 
around a central massive black hole (BH). However, the use of a wide slit
affects spectra of any objects with steep velocity gradients (\eg shocks 
or contact discontinuities).

\begin{figure*}
\includegraphics{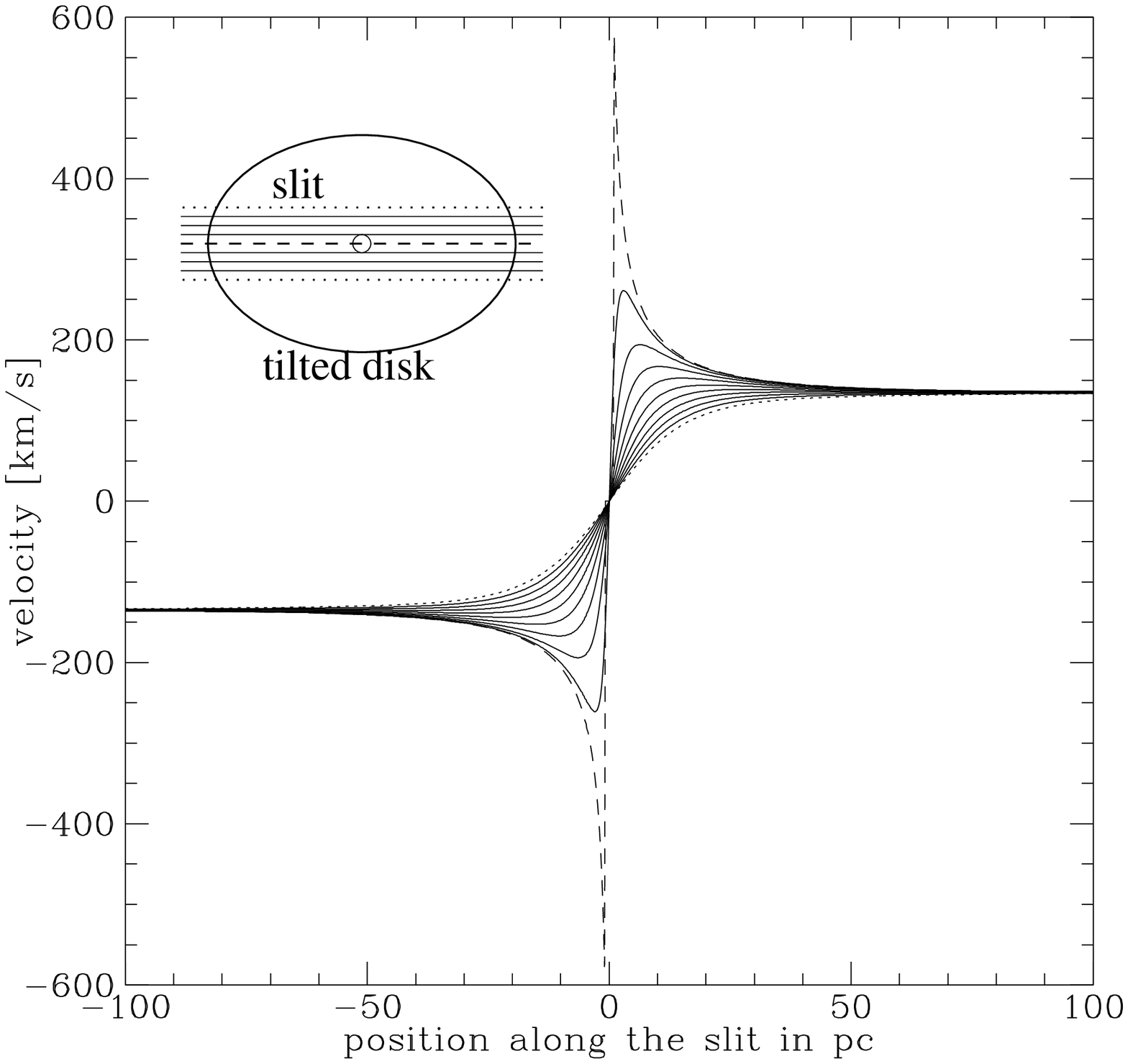}
\includegraphics{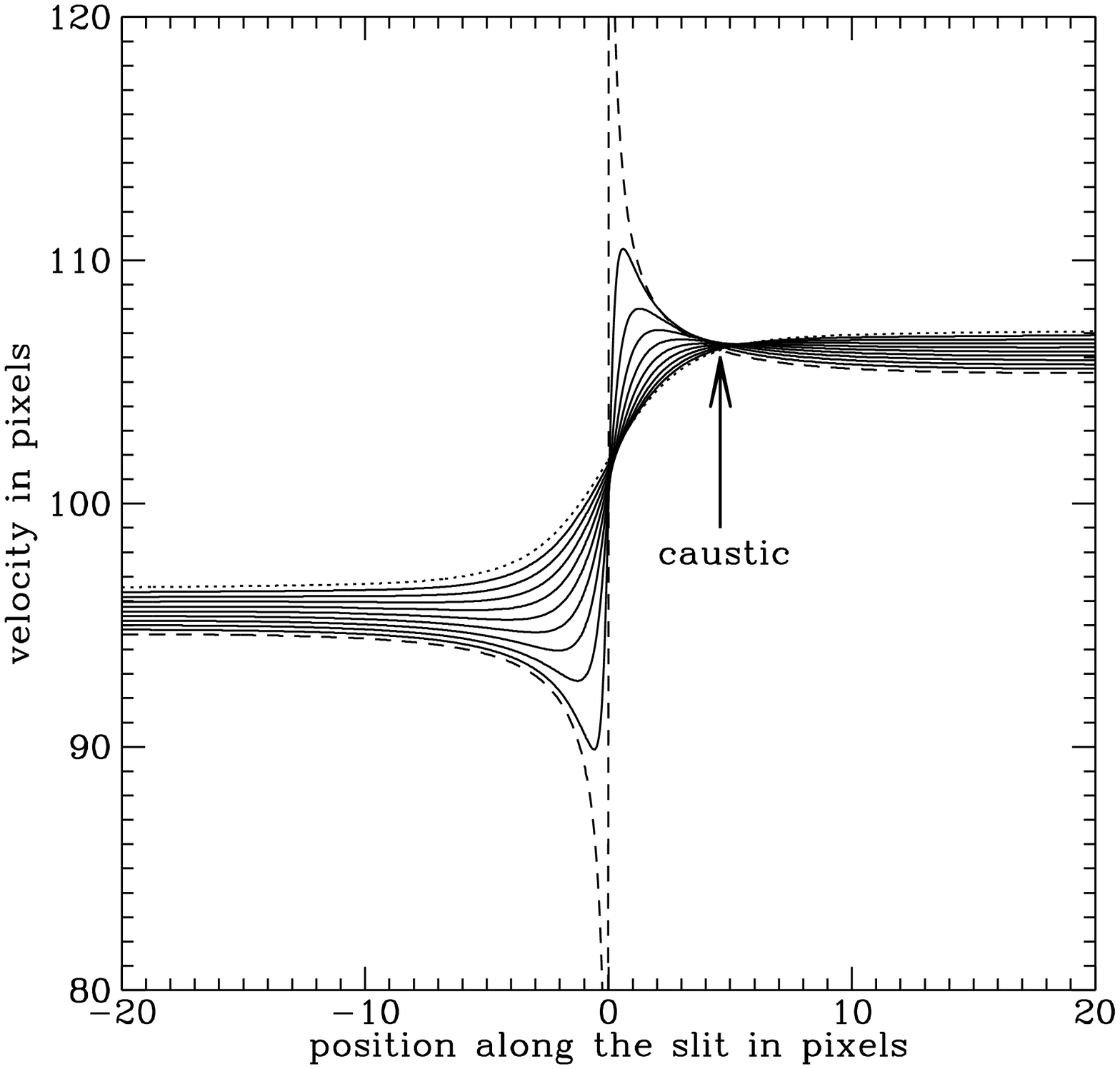}
\includegraphics{maciejewskiw3.ps}
\vspace{4cm}
\caption{{\it Left:} 
Position-velocity diagram for the disk inclined at 60\deg,
rotating in a potential of a $10^8$ \solm BH plus extended density
distribution $\sim R^{-1.8}$. Velocities are sampled along cuts parallel to 
the line of nodes, shown in the inset at upper left.
{\it Center:} The light pattern on the spectrograph's detector for G750M 
grating on STIS (only the top half of the slit is sampled).
{\it Right:} The light from entire slit integrated within the detector's 
pixels.}
\end{figure*} 

\section{A slit wider than the PSF}
The conventional approach to the determination of the BH mass from
emission-line spectra involves fitting a PSF-convolved 
circular-speed curve to the observations well into the nucleus of the
galaxy, where the BH-related velocity rise inwards becomes apparent.
A wide slit samples off-center velocities (solid lines in the left
panel of Fig.1) in addition to the central rotation curve (the dashed 
line). When the disk is observed in some emission line through a long-slit
spectrograph, the pattern of intensity resembles the left panel of Fig.1, but
it is modified by the geometry of the spectrograph's optics: the diffraction
pattern produced by light that enters near one edge of the slit (dotted line)
is displaced with respect to light of the same frequency that enters at the 
corresponding point on the other edge of the slit. Thus
the difference in position across the slit is 
seen on the detector as an {\it instrumental velocity offset}. To reflect 
the light pattern on the detector, the left panel of Fig.1
has to be modified by shifting each sample line in the dispersion direction 
by a constant amount, different for each line, because each line samples
a given position across the slit. 

The result for one half of the slit is presented in the central panel of 
Fig.1 (a plot for the other half is point-symmetric to this one). We see
that in the presence of a velocity gradient across the slit, the instrumental
velocity offset competes with the Doppler shifts: this 
offset wins at large radii, and at small radii the Doppler shifts take over.
In between, both factors are of similar strength: light 
from all positions across the slit converges at one effective 
velocity, forming {\it the caustic}. At radii interior to the caustic,
two maxima in the light distribution are present (Fig.1, right): one 
from the slit center (showing Keplerian rise), and one from the slit 
edges (passing through zero velocity at the nucleus). Thus the 
position-velocity diagrams for rotating disks are rich in structure;
the information contained in them is lost in the conventional fitting of 
a Gaussian emission line profile, because this reduces them to a single 
position-velocity line. Below we explore the entire 2-dimensional light 
distribution.

\section{A new BH mass estimator}
The position of the caustic along the slit is quite independent of the 
stellar density profile, and it can indicate the presence, and betray the 
mass $M_\bullet$, of the BH. A wide slit contains many narrow slits 
within it, and it can provide information that otherwise would
need two offset thin slits. If the caustic occurs at a position 
$\alpha$ down the $2\delta$-wide slit, then velocities at the slit 
center ($v_c$) and at the slit edge ($v_e$) are related by
$ D v_c(\alpha) = D v_e(\alpha) + B\delta $, where $B$ is
the plate scale in the dispersion direction, and $D$ is the spectrograph's 
dispersion. If the only unknowns in the problem are the disk's inclination 
angle $i$ and $M_\bullet$, then this additional constraint allows us to 
recover $i$ and $M_\bullet$ independently, rather than $M_\bullet \sin i$.
In the case of the slit placed along the line of nodes, the solution
takes a particularly simple form:
\[
M_{\bullet} = \frac{\alpha d}{G} \left(\frac{y}{D \sin i} \right)^2 ,
\hspace{2cm}
\cos i= \frac{\delta}{\alpha} \left( \left(1 - \frac{B\delta}{y}\right)^{-4/3} -
 1 \right)^{-1/2},
\]
where $y$ is the position of the caustic in the dispersion direction, $d$
is the distance to the galaxy, and $G$ is the gravitational constant.

In addition to recovering $i$ and $M_\bullet$ independently, this method
gives higher sensitivity to BH detection than traditional methods based 
on the Keplerian rise in velocity occurring inside the BH's sphere of influence,
because it exploits an artifact at the outer edge of this sphere. For the
same reason though, it only yields the mass within the radius of the 
caustic, and it cannot determine whether
the measured mass comes from the BH. On the other hand, the very existence
of the caustic comes from the steep velocity rise towards the nucleus,
which itself is characteristic of a massive BH. Thus the main advantage of
our technique is in detecting BHs in galaxies for which we cannot
achieve the resolution required to follow the Keplerian rise in velocities
inwards.

\section{Inclined slits}
Since nuclear disks frequently lie in a plane that differs
markedly from that of galaxy's main disk, one usually does not know the
line of nodes before nuclear spectra are taken. When the slit lies along 
the line of nodes (like in Fig.1), cuts offset by $\pm\delta$ below and 
above the nucleus map to the same line in the position-velocity plot. The
instrumental velocity offset shifts these lines up and down, and
generates two lines: $y=Dv+B\delta$ and $y=Dv-B\delta$. For the disk
with an opposite sense of rotation, the two cuts are mapped into two other 
lines: 
$y=D(-v)+B\delta = -(Dv-B\delta)$, and $y=D(-v)-B\delta = -(Dv+B\delta)$,
which are just mirror symmetries of the previous two. 
If the slit is not placed along the line of nodes, then equidistant cuts 
passing below and above the nucleus sample two different velocity
fields: $v(\alpha) = -v(-\alpha)$, and it matters which is moved up and 
which down. Thus the long-slit spectrum of a rotating disk taken with
a wide slit depends qualitatively on the sense of rotation of the disk.
For the same disk, one spectrum displays the caustic clearer that the 
other, therefore in our method the detectability of a BH with a given 
spectrograph setup depends on the sense of rotation of its accretion disk.

In Fig.1, where the slit is placed along the line of nodes, the outer 
envelope of the light distribution near the nucleus is formed by light 
coming through the slit center. This is not so for inclined
slits (Fig.2, left panel): there the outer envelope comes from the sides 
of the slit -- the dashed line associated with the slit center has a much 
narrower peak. Consequently, fitting profiles from an infinitely thin slit 
to the observed light distribution would result in a considerable 
overestimation of the BH mass when the traditional method is being used.

\begin{figure*}
\includegraphics{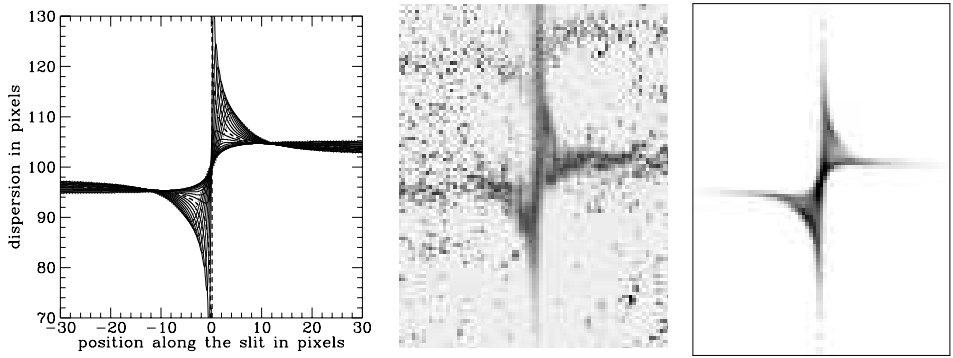}
\includegraphics{maciejewskiw5.ps}
\includegraphics{maciejewskiw6.ps}
\vspace{4cm}
\caption{{\it Left:} Line sampling of our model of light distribution in 
the spectrum of a nuclear disk around a BH with parameters that we derived 
for M84 (the nuclear cut through the slit center is dashed). 
{\it Middle:} STIS $H\alpha$ emission-line spectrum of M84 
(Bower et al.\ 1998). The dispersion direction is vertical. {\it Right:}
light distribution in our model spectrum of M84 convolved with the PSF,
and integrated over detector pixels (slit is offset by 0.03 arcsec from 
the nucleus).}
\end{figure*} 

\section{Confronting observations. Conclusions.}
The discussion above has been confined to the distribution of the principal
maxima in the diffraction pattern produced on the spectrograph's detector.
For our final models, we convolved these diagrams with the instrumental
PSF on STIS approximated by a sum of Gaussians. Convolution was done by
adding in intensity contributions from Airy disks formed on the slit and
truncated by it. 

We confronted then our predicted light pattern with the most detailed  
long-slit spectrum of nuclear emission so far (M84, Bower \etal, 1998, 
{\it ApJ}, 492, L111): the last one shows two light maxima at 
radii close to the nucleus (Fig.2, middle panel). 
Although they were interpreted as coming
from two physically distinct nuclear components, the observed light 
pattern has the same structure as our models in Fig.1. We 
interpret the point where the track 
of maximum light splits into two as the caustic, from which we estimate 
the disk to be inclined at 74\deg, and the BH mass to be
$4 \times 10^8$ \solm, smaller than that of Bower~\etal by a factor of 4.
This confirms our finding that BH masses derived by the traditional 
method may be overestimates. The left-right asymmetry of the observed 
spectrum comes from the slit not being centered on the nucleus: the 
right panel of Fig.2 shows our predicted light pattern for the slit
offset by .03 arcsec from the nucleus. Thus all features in the nuclear
spectrum of M84 can be explained as coming from a thin disk in circular 
motion around the central BH.

The finite width of the slit generates complex patterns in the spectra
that can have considerable diagnostic power if they are modeled with 
adequate sophistication. They allowed us to develop a new method for 
estimating the BH mass that gives higher 
sensitivity to BH detection than traditional methods.

\end{document}